# Tracking photophysical relaxation in spiropyran with simulated time-resolved X-ray absorption spectroscopy

Check for updates

Lina Fransén ①[1] ✉, Benjamin Roux[1], Sonia Coriani ①[2], Saikat Nandi ①[3] & Morgane Vacher ①[1] ✉

Spiropyran is a prototype photoswitch that undergoes photoinduced ring-opening via C-O bond fission. The quantum yield for this photochemical reaction is low, with photophysical relaxation being the dominant process. Previous theoretical studies have suggested that photophysical relaxation proceeds via cleavage and subsequent reformation of a C-N bond. However, experimental evidence for this mechanism is lacking; most time-resolved studies have probed in the ultraviolet-visible domain, where C-N bond fission is unlikely to leave clear signatures. Here, we use non-adiabatic dynamics simulations in conjunction with X-ray absorption spectroscopic calculations to investigate the possibility of tracking the photophysical relaxation in spiropyran with time-resolved X-ray absorption spectroscopy at the nitrogen K-edge. The simulations predict that cleavage of the C-N bond results in a transient red-shift of the X-ray absorption. These results indicate a potential experimental route to gain a mechanistic understanding of the efficient photophysical relaxation that limits the efficiency of spiropyran-based photochromic systems.

Spiropyrans are widely studied photochromic compounds that find application in areas including biological imaging[1] and memory storage[2]. Their utility stems from their photoswitching ability—namely, their capacity to undergo reversible, light-induced changes in molecular structure and associated properties, such as optical absorption. The structural transformation, known as photoisomerization, involves conversion of the closed-form spiropyran into an open merocyanine form via ultraviolet (UV)-induced cleavage of the $C_{spiro}$–O bond (Fig. 1). In the spiropyran form, the π-systems of the chromene (oxygen-containing) and indoline (nitrogen-containing) moieties are orthogonal and largely non-interacting. Upon isomerization to the merocyanine form, these π-systems become conjugated, resulting in a red-shifted absorption: merocyanine absorbs in the visible spectral region, whereas spiropyran absorbs in the UV.

Owing to their numerous applications, spiropyrans have been studied extensively experimentally using transient absorption spectroscopy, with most studies probing in the UV-visible[3–7] and infrared (IR)[8,9] domains. For example, Ernsting and Arthen-Engeland obtained a merocyanine formation time constant of 0.9 ps from the rise time of the merocyanine ground state absorption band in the visible region[5]. Rini et al. reported, using femtosecond UV-mid-IR pump-probe spectroscopy, a considerably longer merocyanine formation time of 28 ps[8]. They found, moreover, that the major relaxation channel following light absorption is not merocyanine formation but internal conversion to the $S_0$ state of spiropyran, with a quantum yield of ≥90%. In line with this, Kohl-Landgraf et al. reported a merocyanine formation yield of only 3.3% for a water-soluble spiropyran[6]. Fidder et al. proposed, based on a weaker energy gap dependence than expected from the energy gap law, that large conformational changes are responsible for the efficient photophysical $S_1 \rightarrow S_0$ internal conversion in spiropyran[9]. The nature of these conformational changes, however, could not be discerned.

A series of computational studies on spiropyrans have provided insight into the mechanism of photochromic ring-opening via $C_{spiro}$–O bond fission[10–16] and suggested that photophysical relaxation proceeds through breakage and subsequent reformation of the $C_{spiro}$–N bond[11–13,15,16]. In an early computational study, Celani et al. investigated chromene as a simplified spiropyran model using CASSCF static electronic structure calculations[10]. Chromene was chosen as a simplified spiropyran model as the lowest bright state ($S_1$) of unsubstituted spiropyran is located on the chromene moiety, and previous studies had shown that only this moiety exhibits photochromism. Following $S_1$ excitation, the wave packet was predicted to evolve toward ground-state merocyanine through an $S_1/S_0$ conical intersection (CI) reached through $C_{spiro}$–O bond extension. Building on this work, Sanchez-Lozano et al. investigated a less simplified spiropyran model using similar computational methods[11]. In this model, the indoline unit was retained but simplified by removing the fused benzene ring and replacing the three methyl groups with hydrogen atoms. Retaining the indoline moiety allowed the identification of a second $S_1/S_0$ CI, located along the $C_{spiro}$–N bond extension. The $C_{spiro}$–N bond was proposed to reform on the $S_0$ state. Decay through this CI was thus suggested to be photophysical, leading to the reformation of spiropyran in its ground state. Subsequent static electronic structure calculations by Liu and Morokuma[12,13] and on-the-fly non-adiabatic dynamics studies by Zhang et al.[14] and Granucci and Padula[15] on a yet less drastically simplified spiropyran model —where the only simplification was the substitution of three methyl groups with hydrogen atoms—further supported this mechanistic picture: merocyanine formation via $S_1 \to S_0$ relaxation through a CI located along the $C_{spiro}$–O bond extension, competing with photophysical relaxation through a CI along the $C_{spiro}$–N bond extension. However, neither Granucci and Padula nor Zhang et al. obtained a significant amount of $C_{spiro}$–N bond reformation in their non-adiabatic dynamics simulations; as far as we are aware, this has only been observed in dynamics simulations on the related compound spironaphthopyran[7].

[1]Nantes Université, CNRS, CEISAM, UMR 6230, F-44000 Nantes, France. [2]Department of Chemistry, Technical University of Denmark, Kemitorvet Building 207, DK-2800, Kongens Lyngby, Lyngby, Denmark. [3]Université de Lyon, Université Claude Bernard Lyon 1, CNRS, Institut Lumière Matière, 69622, Villeurbanne, Lyon, France.  ✉e-mail: lina.fransen@univ-nantes.fr; morgane.vacher@univ-nantes.fr

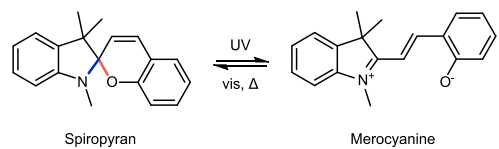

**Fig. 1 | Photochromic conversion between spiropyran and merocyanine.** The $C_{spiro}$–O bond that breaks during merocyanine formation is highlighted in red. The $C_{spiro}$–N bond that has been theoretically proposed to break transiently in a competing photophysical relaxation channel is highlighted in blue.

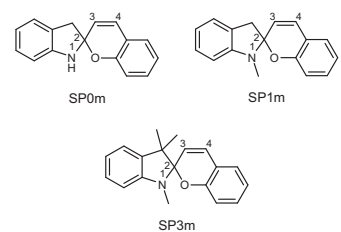

**Fig. 2 | Methyl-hydrogen substituted spiropyran models SP0m and SP1m, along with the parent molecule SP3m.** The SP1m model is used in the present work.

To our knowledge, the $C_{spiro}$–N bond fission mechanism of photophysical relaxation, first predicted theoretically over a decade ago[11], has not yet been experimentally verified or tested. Breakage of the $C_{spiro}$–N bond is unlikely to leave clear signatures in the UV-visible region, where most pump-probe experiments have been conducted. A powerful experimental method for probing structural and electronic motions in an element-specific manner during excited-state dynamics is time-resolved X-ray absorption spectroscopy (TRXAS)[17–22]. This technique has been shown to be sensitive to bond length changes[23–25].

In the present work, we use on-the-fly mixed quantum-classical non-adiabatic dynamics simulations in conjunction with X-ray spectroscopic calculations to propose that TRXAS at the nitrogen K-edge has the potential to detect $C_{spiro}$–N bond breakage and reformation. A deeper understanding of the mechanism responsible for the efficient photophysical relaxation in spiropyran could aid in the design of more efficient photochromic systems.

## Methods

The time evolution of the coupled electron-nuclear dynamics of spiropyran was obtained through non-adiabatic dynamics simulations with the surface hopping approach coupled with time-dependent density functional theory (TD-DFT) for the electronic structure. Nitrogen K-edge TRXAS spectra were calculated along representative surface hopping trajectories using the restricted active space (RAS) method. Similar protocols for simulating TRXAS spectra have been employed previously; however, most studies have investigated considerably smaller systems[26–31]. A simplified spiropyran model was employed throughout to reduce the computational cost; a discussion of the model choice is provided in the next subsection. The two subsequent subsections outline the computational details for the surface hopping simulations and the TRXAS calculations, respectively.

## Model system

As outlined in the Introduction, many computational studies on spiropyrans have employed simplified model systems to reduce computational costs. Among these, the least drastic simplification, used in both recent static studies[12,13] and non-adiabatic dynamics studies[14–16], involves replacing the three methyl groups with hydrogen atoms. This model, in which all three methyl groups are replaced, is referred to herein as SP0m, while the parent compound is labeled SP3m (Fig. 2).

Using surface hopping simulations coupled with semiempirical OM2/MRCI electronic structure calculations, Zhang et al. demonstrated that the excited-state dynamics of SP0m closely resemble those of SP3m, with the main distinction being an artificial introduction of proton transfer from the nitrogen atom or from a carbon atom to the oxygen atom in the model compound[14]. Given this overall similarity, the present work also adopts a methyl-hydrogen substituted spiropyran model. However, since methyl-hydrogen substitution on the nitrogen atom can influence reactivity, as demonstrated in the reference mentioned above, we retain the methyl group attached to the nitrogen while substituting the other two methyl groups. To our knowledge, the present work is the first to consider this specific spiropyran model, which we designate SP1m (Fig. 2). The use of SP1m is supported by the similarity of its geometric parameters, excitation energies, and $S_1$ natural transition orbitals at the Franck-Condon (FC) point compared to those of SP3m (see Supplementary Note 1).

### Non-adiabatic dynamics simulations

**Electronic structure method.** The electronic structure method used to compute the potential energy surfaces and their couplings for the surface hopping simulations in this work was TD-DFT with the Tamm-Dancoff approximation (TDA) and the $\omega$B97XD[32] functional. TD-DFT with TDA and hybrid functionals has been demonstrated as an efficient alternative to multireference methods when the latter are computationally prohibitive due to the system size[33]. The $\omega$B97XD functional was selected as it provides $S_1$ and $S_0$ potential energies that are in good agreement with previously reported MS-CASPT2 reference data[12] for SP0m at key geometries likely to be explored during the dynamics, including the FC point, CIs along $C_{spiro}$–O and $C_{spiro}$–N bond extensions, and initially formed merocyanine conformers (see Supplementary Note 2.1). TD-DFT with the $\omega$B97XD functional has, moreover, previously been found to provide excitation energies and oscillator strengths at the FC point in good agreement with reference RI-CC2 data[34].

The 6-31G* basis set was used for all (TD-)DFT calculations, which were carried out with Gaussian 16, rev. A03[35]. The *tight* SCF convergence criteria and *fine* integration grid were requested.

**Propagation of the coupled electron-nuclear dynamics.** Mixed quantum-classical dynamics simulations were carried out using the surface hopping method as implemented in the SHARC software (version 3.0)[36,37]. Initial conditions were sampled from two separate 0 K Wigner distributions generated from the ground-state frequencies of SP1mt and SP1mc (two ground-state spiropyran conformers, see the first subsection in the Results and Discussion section). A total of 200 trajectories were propagated, 100 for each of SP1mt and SP1mc. All trajectories were initiated on the $S_1$ state. A nuclear time step of 0.5 fs was used; satisfactory energy conservation with this time step is demonstrated in Supplementary Fig. 4. The electronic wave function was propagated using the local diabatization method[38] with 100 substeps, together with the energy-based decoherence correction with a decay factor of 0.1 $E_h$[39]. Three electronic states, $S_0$–$S_2$, were considered in the non-adiabatic dynamics simulations. Hopping probabilities between $S_1$ and $S_2$ were computed from the time evolution of the electronic amplitudes. A well-documented deficiency of TD-DFT is its poor description of CIs with the electronic ground state[40]. Therefore, hops to $S_0$ were forced when the energy gap between the active state and $S_0$ fell below 0.1 eV. The nuclear velocity vectors were rescaled isotropically to conserve the total energy after a hop. The trajectory analysis was terminated if the total energy at a given time step deviated by more than 0.5 eV from the initial value. This condition was met for 3 trajectories (1.5%), all of which had previously undergone $C_{spiro}$–N bond fission. The effect of decreasing the threshold to 0.2 eV is discussed in Supplementary Note 2.3. Additionally, 34 trajectories (17%) were terminated prematurely due to a failure of convergence in the electronic structure calculations, despite increasing the maximum number of SCF cycles from the default of 128 to 1024. Out of these, 32 occurred after $C_{spiro}$–N bond fission, and 2 followed $C_{spiro}$–O bond fission. When the $C_{spiro}$–N and $C_{spiro}$–O bonds are stretched, the $S_1/S_0$ energy gap is reduced; similar issues with electronic structure failures in regions of near-degeneracy with the $S_0$ state have previously been reported for single-reference electronic structure methods[33]. All trajectories are considered in the analysis up until the time point of eventual failure (i.e., up until the time step when energy conservation or convergence problems were encountered). Since all failures occurred after $C_{spiro}$–O or $C_{spiro}$–N bond fission, they do not impact the calculated branching ratios of relaxation via $C_{spiro}$–O versus $C_{spiro}$–N bond fissions. However, we acknowledge that the ultimate fate of the affected trajectories (i.e., whether they would remain ring-open or if the bond would reform) cannot be known. Further details on the prematurely terminated trajectories are provided in Supplementary Note 2.3.

The $C_{spiro}$–N and $C_{spiro}$–O bonds were considered broken when they were extended to 2.1 Å. A reduction of this arbitrary threshold to 1.8 Å results in the reclassification of one trajectory from non-reactive to having undergone $C_{spiro}$–O bond fission. This trajectory appears to begin to undergo $C_{spiro}$–O bond fission near the end of the simulation.

### TRXAS calculations

Nitrogen K-edge XAS spectra were calculated with the RAS approach implemented in the OpenMolcas[41,42] software (version 22.10-354-g7f2c128[43]). The ANO-RCC-VDZP[44] basis set and the atomic compact Cholesky decomposition[45] were used. The N 1s orbital was placed in the RAS1 space, where a maximum of one hole was allowed. The RAS2 space, analogous to the active space in CASSCF, comprised 12 electrons in 11 orbitals. These include five $\pi$ orbitals (three on the chromene subunit and two on the indoline subunit), four $\pi^*$ orbitals (two on each subunit), and a pair of $\sigma/\sigma^*$ orbitals. This active space can be labeled systematically as RAS(14,1,0;1,11,0), using the notation RAS(n,l,m;i,j,k), where $i$, $j$ and $k$ are the number of orbitals in RAS1, RAS2 and RAS3, $n$ is the total number of electrons, $l$ is the maximum number of holes in RAS1, and $m$ the maximum number of electrons in RAS3.

The active space used for the TRXAS calculations must include the $\sigma$ and $\sigma^*$ orbitals required for an adequate description of structures exhibiting broken $C_{spiro}$–O and $C_{spiro}$–N bonds, along with the $\pi$ and $\pi^*$ orbitals necessary for a satisfactory description of both valence- and core-excited states. To meet these requirements, distinct RAS2 active spaces were used to compute XAS spectra along trajectories undergoing $C_{spiro}$–O and $C_{spiro}$–N bond breakages. Specifically, the $C_{spiro}$–N $\sigma/\sigma^*$ orbital pair was included for trajectories experiencing $C_{spiro}$–N bond fission, while the $C_{spiro}$–O $\sigma$ and $\sigma^*$ orbitals were included for trajectories undergoing fission of this bond. In the following, the two active spaces are referred to as $RAS_{CN}$(14,1,0;1,11,0) and $RAS_{CO}$(14,1,0;1,11,0). This approach builds on the work by Liu and Morokuma, who employed different CASSCF active spaces to compute valence potential energy profiles for $C_{spiro}$–O and $C_{spiro}$–N bond fissions[12].

The active spaces, which are depicted as isodensity plots in Supplementary Note 3.1, were validated with respect to two aspects. Firstly, Supplementary Fig. 10 shows that the valence potential energies computed using $RAS_{CN}$(14,1,0;1,11,0) and $RAS_{CO}$(14,1,0;1,11,0) along representative surface hopping trajectories undergoing $C_{spiro}$–N or $C_{spiro}$–O bond fission are in good qualitative agreement with the potential energy curves provided by TDA-$\omega$B97XD. This suggests that the two active spaces provide reliable descriptions of bond breakage. Secondly, Supplementary Fig. 11 shows that the XAS spectra calculated using $RAS_{CN}$(14,1,0;1,11,0) and $RAS_{CO}$(14,1,0;1,11,0) at the FCpoints are nearly identical. This suggests that selecting one active space over the other does not introduce bias.

TRXAS spectra were calculated every 10 fs along four representative surface hopping trajectories, comprising two SP1mt and two SP1mc trajectories. For each conformer, one trajectory undergoing $C_{spiro}$–N bond fission and another undergoing $C_{spiro}$–O bond fission were selected. The nuclear geometries undergo rapid changes during the breaking of the $C_{spiro}$–N or $C_{spiro}$–O bond. To promote the conservation of the RAS2 spaces throughout these changes, the valence-excited states were computed sequentially using RASSCF every femtosecond (i.e., every other time step from the dynamics simulations). Three valence-excited states ($S_0$, $S_1$, and $S_2$) were included and treated using the state-average approach. Core-excited states were computed every 10 fs with a state average of 40, using the orbitals for the valence-excited states as initial guesses. The core-excited states were reached with the HEXS approach[46], and the *supym* keyword was activated for the N 1s orbital to prevent it from rotating out of the RAS1 space. The restricted active space state interaction (RASSI) method was used to compute excitation energies and oscillator strengths between the three valence-excited states and the 40 core-excited states every 10 fs. The resulting stick spectra were broadened with a Voigt function with a Lorentzian full width at half maximum (FWHM) of 0.11 eV[26,47] that accounts for the core-hole lifetime broadening and a Gaussian standard deviation of 0.2 eV that accounts for experimental resolution.

To assess the robustness of the spectral feature characteristic of $C_{spiro}$–N bond fission across different computational methods, XAS spectra were computed at three geometries along a representative SP1mt trajectory exhibiting $C_{spiro}$–N bond fission with three methods in addition to RASSCF: (i) RASPT2, (ii) equation-of-motion coupled-cluster (EOM-CCSD), and (iii) TDA and the (initial) maximum overlap method (IMOM)[48]. The RASPT2 calculations were performed with an imaginary shift of 0.1 $E_h$. The CCSD core-excited states were reached with the fc-CVS-EOM-CCSD[49] approach. For the (IMOM-)TDA calculations, the $\omega$B97XD functional was used. The CCSD and (IMOM-)TDA XAS calculations were carried out with the 6-31$^+$G* basis set using Q-Chem (v. 6.1)[50].

## Results and discussion

The results are now presented, starting with a description of two ground state spiropyran conformers and the character of the lowest-lying valence-excited electronic state at these geometries. Next, the photodynamics of spiropyran, as described by the surface hopping simulations, are discussed. Finally, the predicted capabilities of TRXAS for providing fingerprints of the photophysical relaxation in spiropyran are explored.

### $S_0$ and $S_1$ states at the Franck-Condon point

Two ground state conformers of SP1m, differing by the N1-C2-C3-C4 dihedral angle $\alpha$ (see Fig. 2 for atom numbering), were optimized at the $\omega$B97XD level. The two conformers are in the following referred to as SP1mt ($\alpha = -144°$) and SP1mc ($\alpha = -103°$). Analogous structures were obtained at the same level of theory for the parent compound SP3m (Supplementary Tables 1 and 2) and in previous studies also for SP0m at the CASSCF[12] and OM2/MRCI[14] levels of theory. This establishes the robustness of the presence of two minima across different model systems and electronic structure methods. We note, however, that only one ground state structure was obtained for SP0m by Granucci and Padula at the semiempirical rAM1/FOMO-CI(10,7) level[15]. SP1mt and SP1mc are essentially isoenergetic at the $\omega$B97XD level ($\Delta E = 0.04$ eV), and they are therefore assumed to both be present at room temperature. Both forms are therefore considered in the non-adiabatic dynamics simulations and TRXAS calculations discussed in the next subsections.

The $S_1$ states at the optimized SP1mt and SP1mc geometries correspond to $\pi \to \pi^*$ transitions localized on the chromene subunit (Fig. 3), in agreement with several previous theoretical investigations[14,15]. Further, a lowest bright state localized on the chromene subunit is consistent with the experimental assignment made by Tyer and Becker who, based on a comparison of the spectrum of SP3m with those of indoline and 2,2-diethyl-chromene, concluded that the chromene moiety is responsible for the lowest-energy region of the spectrum[51]. The surface hopping trajectories discussed in the next subsection were initiated on the $S_1$ state.

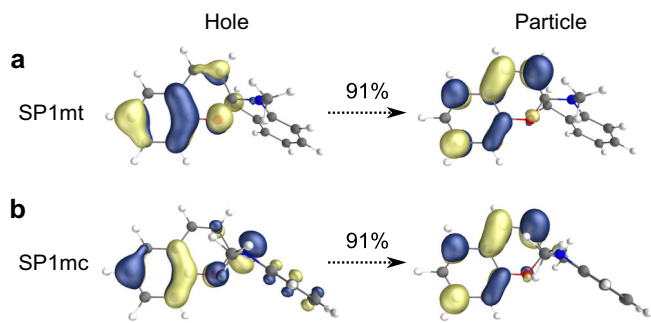

**Fig. 3 | Natural transition orbitals for the $S_1$ state.** Panels (**a**, **b**) show the orbitals for the SP1mt and SP1mc conformers, respectively. The weight associated with each particle-hole pair is displayed above the arrow. The orbitals were computed at the TDA-ωB97XD level of theory and rendered using an isodensity value of 0.05.

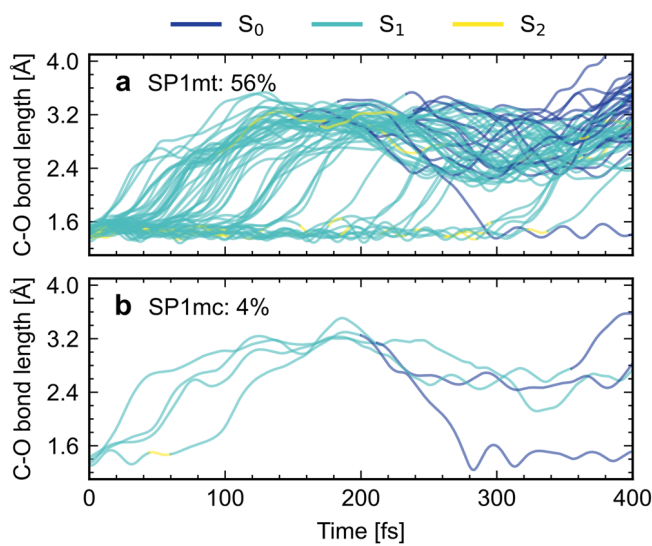

**Fig. 4 | Time evolution of the $C_{spiro}$–O bond lengths for the trajectories that undergo fission of this bond. a** Shows the trajectories initiated from the SP1mt ground state conformer, and (**b**) shows those initiated around SP1mc. The percentage of trajectories for each conformer that undergo $C_{spiro}$–O bond breakage is indicated. The trajectories are color-coded according to the active electronic state.

**Coupled electron-nuclear dynamics**
In the 400 fs following excitation to the $S_1$ state, 98.5% of all trajectories undergo fission of either the $C_{spiro}$–O or the $C_{spiro}$–N bond. We begin by discussing the well-studied photochemical route, where the $C_{spiro}$-O bond is cleaved to form merocyanine, and then proceed to the competing photophysical relaxation channel, which is the focus of the present work. An overview of the photodynamics is thereafter provided, before presenting TRXAS calculations along representative surface hopping trajectories in the next subsection.

Figure 4 shows the time evolution of the $C_{spiro}$–O bond lengths for the trajectories that exhibit fission of this bond. The trajectories are separated based on the initial ground state conformer; the branching ratios and reaction times differ considerably between SP1mt and SP1mc, as will be discussed in more detail below. Consistent with the established mechanism, $C_{spiro}$–O bond extension results in a reduction of the energy gap between the $S_1$ and $S_0$ states. During the 400 fs simulation, 53% of the trajectories that exhibit $C_{spiro}$–O bond breakage undergo $S_1 \rightarrow S_0$ internal conversion, as indicated by the color coding of the trajectories in Fig. 4.

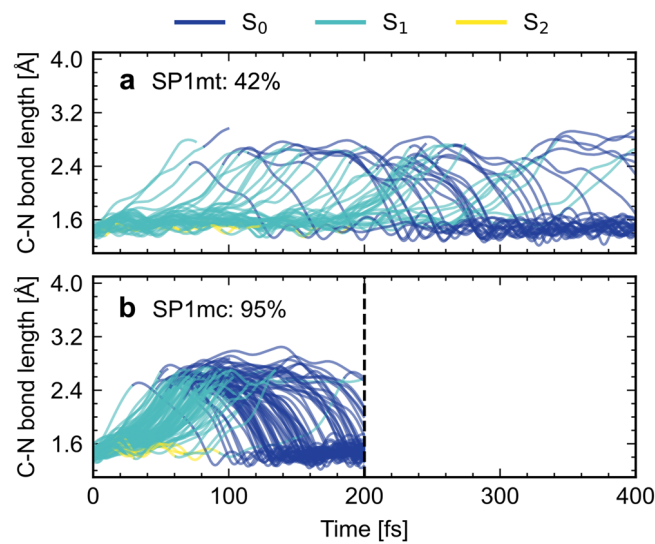

**Fig. 5 | Time evolution of the $C_{spiro}$–N bond lengths for the trajectories that undergo fission of this bond. a** Shows the trajectories initiated from the SP1mt ground state conformer, and (**b**) shows those initiated around SP1mc. The percentage of trajectories for each conformer that undergo $C_{spiro}$–N bond breakage is indicated. The SP1mc trajectories undergoing $C_{spiro}$–N bond fission were propagated for only 200 fs as they reacted quickly, while the remaining trajectories were propagated for 400 fs. The trajectories are color-coded according to the active electronic state.

The bulk of the trajectories stay $C_{spiro}$–O ring-opened throughout the simulation time and start to evolve toward the formation of more stable merocyanine forms through rotation around one or several methine bridge bonds. For a small fraction of the trajectories—3% of those undergoing $C_{spiro}$–O bond breakage—the bond is reformed after internal conversion to $S_0$, leading to the regeneration of spiropyran in its ground state.

Liu and Morokuma proposed that hydrogen out-of-plane (HOOP) motion is required for efficient internal conversion along the $C_{spiro}$–O bond fission channel, as MS-CASPT2 provides a significant $S_1/S_0$ energy gap at the CASSCF-optimized CI located along the stretching of the $C_{spiro}$–O bond[12]. Subsequent non-adiabatic dynamics studies have both supported[14,52] and challenged[15] this proposed HOOP mechanism. The TDA-ωB97XD method used for the non-adiabatic dynamics simulations herein provides a finite $S_1/S_0$ energy gap comparable to that provided by MS-CASPT2 at the CASSCF-optimized CI for SP0m (Supplementary Fig. 2). However, no evidence for HOOP-assisted $S_1 \rightarrow S_0$ internal conversion is found in the present work (Supplementary Note 2.4). As pointed out by Gonon et al., the CI is not necessarily absent but may simply be displaced at the CASPT2 level[53] (and, similarly, at the TDA-ωB97XD level used herein).

Moving on to the competing photophysical relaxation channel, Fig. 5 shows the trajectories that undergo cleavage of the $C_{spiro}$–N bond. Analogous to $C_{spiro}$–O bond dissociation, the extension of the $C_{spiro}$–N bond reduces the $S_1/S_0$ energy gap. This leads to $S_1 \rightarrow S_0$ internal conversion for 85% of the $C_{spiro}$–N bond-breaking trajectories during the simulation time. Of the trajectories that do not hop to $S_0$, 90% were prematurely terminated due to problems with the electronic structure calculation or conservation of energy (Supplementary Note 2.3). The $S_1 \rightarrow S_0$ internal conversion occurs at a median $C_{spiro}$–N bond length of 2.5 Å for both SP1mt and SP1mc for the trajectories following the photophysical route, which is considerably shorter than the corresponding median $C_{spiro}$–O bond length of 3.1 Å for the trajectories following the photochemical pathway. This trend is consistent with the CASSCF-calculated potential energy curves for SP0m reported by Liu and Morokuma[12] and with the non-adiabatic dynamics study on SP0m by Granucci and Padula[15].

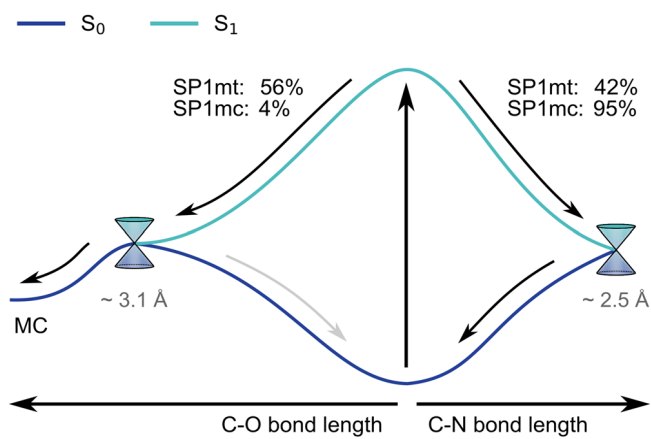

Fig. 6 | Overview of the photodynamics of SP1mt and SP1mc. In the 400 fs following excitation to the $S_1$ state, 98.5% of the trajectories undergo either $C_{spiro}$–O or $C_{spiro}$–N bond dissociation. $C_{spiro}$–O bond fission leads, for the large majority of the trajectories, to merocyanine (MC) formation, while cleavage of the $C_{spiro}$–N bond results in the reformation of spiropyran in its electronic ground state. The numbers below the CIs indicate the median bond lengths at the $S_1 \rightarrow S_0$ hops, which are the same for SP1mt and SP1mc. The potential energy curves are drawn schematically.

Following internal conversion to the $S_0$ state, the $C_{spiro}$–N bond is reformed, leading to the reformation of spiropyran in its electronic ground state. The present dynamics simulations on spiropyran are, to our knowledge, the first to describe a significant amount of $C_{spiro}$–N bond reformation. In the non-adiabatic dynamic study by Zhang et al. on SP0mc, SP0mt, SP3mc, and SP3mt, the yields of $C_{spiro}$–N bond fission were very low (see below for details), and the subsequent reformation of the bond was not discussed[14]. Granucci and Padula, who found only one ground state spiropyran conformer, observed $C_{spiro}$–N bond breakage for approximately one third of their surface hopping trajectories, but a reformation of the bond was not seen[15]. The absence of bond reformation was attributed to the stop conditions: the trajectories were terminated after spending 100 fs in the $S_0$ state.

The photochemistry and photophysics of SP1mt and SP1mc are summarized in Fig. 6. For the SP1mt conformer, the branching ratios of the two bond-breaking pathways are comparable: 56% of the trajectories undergo $C_{spiro}$–O bond breakage and 42% undergo $C_{spiro}$–N bond breakage. The SP1mc conformer, in contrast, exhibits a strong preference for $C_{spiro}$–N bond fission, with 95% of the trajectories favoring this route. Moreover, $C_{spiro}$–N bond breakage occurs in an almost ballistic fashion for SP1mc, while it proceeds in a more gradual manner for SP1mt (Fig. 5). In the non-adiabatic dynamics study by Zhang et al., 0% of the SP0mt trajectories and 0.8% of the SP0mc trajectories underwent $C_{spiro}$–N bond breakage[14]. The corresponding numbers for SP3mt and SP3mc in the same reference are 2.5% and 6.5%, respectively. Although the $C_{spiro}$–N bond fission percentages reported by Zhang et al. are much lower than those observed in the present study, their results hint at the same trend: trajectories starting from the ground-state conformer with a smaller N1-C2-C3-C4 dihedral angle $\alpha$ exhibit a greater propensity for $C_{spiro}$–N bond fission compared to those starting from the conformer with a larger $\alpha$. In light of this, it may be interesting for future work to investigate the influence of the technique used for the sampling of initial conditions on the simulated photochemical outcome for spiropyran. In particular, the quantum thermostat has been proposed to provide more accurate initial conditions for flexible molecules[54,55] than the Wigner distribution used in the present work and in the study by Zhang et al.

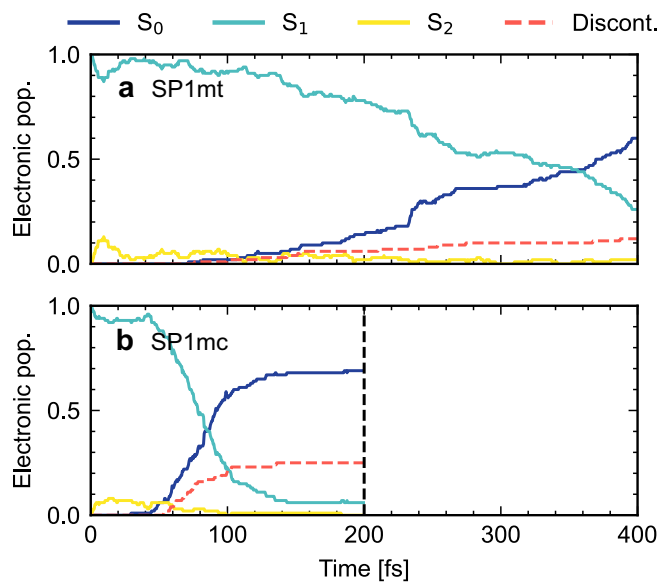

Fig. 7 | Adiabatic electronic populations following excitation to $S_1$. Panel (**a**) shows the results for trajectories initiated from the SP1mt ground state conformer, and panel (**b**) shows those initiated around SP1mc. The populations are shown up 400 fs for SP1mt, but only 200 fs for SP1mc as 95% of the SP1mc trajectories were stopped at 200 fs due to the fast $C_{spiro}$–N bond-breaking reactivity. The fraction of trajectories discontinued due to electronic structure calculation failure or issues with total energy conservation is shown in red.

The different reactivity exhibited by the SP1mt and SP1mc conformers is reflected in the electronic populations, which are displayed in Fig. 7. For the SP1mt conformer, the population in the initially populated $S_1$ state has decayed to 50% of its initial value after ~320 fs, while for the SP1mc conformer, this occurs at ~80 fs. The shorter $S_1$ half-life of the latter conformer can be attributed to the efficient photophysical relaxation via almost ballistic $C_{spiro}$–N bond cleavage (Fig. 5).

The overall quantum yield for merocyanine formation could, in principle, be estimated from the simulation by weighting the yields obtained for the SP1mt and SP1mc conformers using the corresponding ground-state Boltzmann factors, and possibly also accounting for the oscillator strengths for the individual trajectories. However, the Boltzmann factors are very sensitive to the relative energies of the two forms, which in turn are sensitive to the electronic structure method and the spiropyran model (see Supplementary Table 2 and the associated discussion). Assuming arbitrarily an equal amount of both conformers, the present dynamics simulations would predict a merocyanine formation yield of 29%. Both Zhang et al. and Granucci and Padula found in their surface hopping simulations merocyanine formation quantum yields for SP0m that are considerably higher than the experimental upper limit of 10% obtained by Rini et al. for SP3m in tetrachloroethene solution[8]. The authors of both theoretical studies attributed the discrepancy in part to solvent effects. In a recent follow-up study, Granucci and co-workers repeated the surface hopping simulations in three different solvents[16]. The authors noted: "In summary, considering the photoisomerization quantum yield, the agreement with the experimental results is not good. This is due to the presence of a decay channel (the C-N bond breaking) that competes with the C-O cleavage: the determination of $\Phi$ [the quantum yield, authors' note] depends on the quantitative assessment of the relative importance of the two channels, which might in turn depend on details of the $S_1$ potential energy surface"[16]. The sensitivity of the outcome of non-adiabatic simulations on the electronic structure method implied in the quote above has been increasingly recognized in the non-adiabatic dynamics community[33,56,57]. Overall, the simulated quantum yield can thus be expected to be sensitive to the initial conditions, the electronic structure method, and possibly the model system used. However, a precise prediction of the merocyanine formation quantum yield is not essential for the present study, which aims to propose an experiment capable of testing the theoretically predicted mechanism of photophysical relaxation via $C_{spiro}$–N bond breakage and reformation. This is addressed in the next subsection.

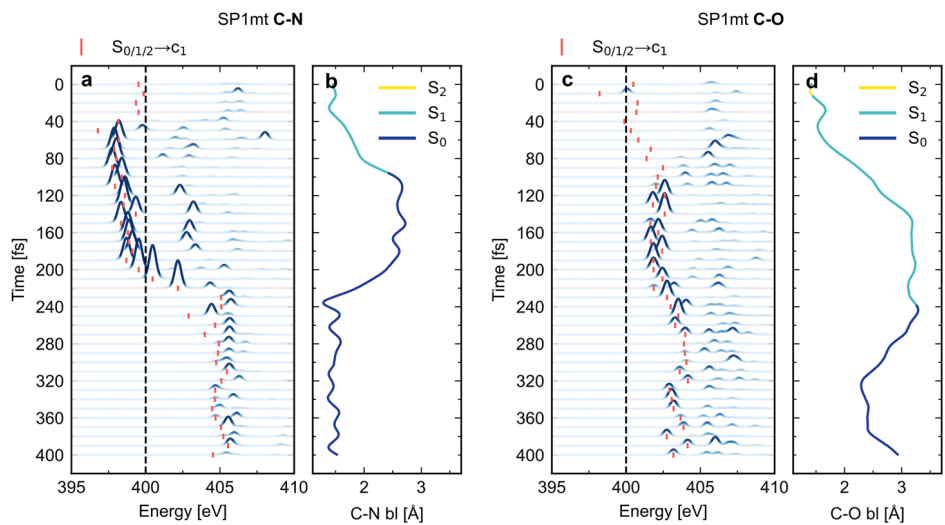

**Fig. 8 | Nitrogen K-edge TRXAS spectra calculated along two representative SP1mt surface hopping trajectories. a, c** Show spectra calculated along trajectories undergoing $C_{spiro}$–N and $C_{spiro}$–O bond cleavage, respectively. The time evolution of the $C_{spiro}$–N and $C_{spiro}$–O bond lengths (bl) along the two trajectories is shown in (**b, d**) respectively. The initial electronic states for the XAS spectra shown in panels (**a, c**) are the active states from the surface hopping simulations, which are indicated by the color coding of the bond lengths in (**b, d**). The red markers in (**a, c**) represent the energy required for excitation from the active electronic state to the lowest-lying core-excited state. No energy shift has been applied to the spectra, which were calculated using RASSCF.

**Tracking photophysical relaxation with TRXAS**

We now turn to investigate the ability of TRXAS at the nitrogen K-edge to provide fingerprints for the photophysical relaxation in spiropyran. Figure 8 shows TRXAS spectra calculated along two representative SP1mt trajectories: one undergoing $C_{spiro}$–N bond breakage and the other undergoing $C_{spiro}$–O bond breakage. TRXAS spectra for two representative SP1mc trajectories are qualitatively similar and are reported in Supplementary Fig. 19. In the case of $C_{spiro}$–N bond fission, the bond cleavage is associated with the emergence of an intense red-shifted signal below 400 eV, which gradually disappears as the bond reforms on the electronic ground state. RASPT2, CCSD, and (IMOM-)TDA all predict similar spectral features (Supplementary Note 3.5). This red-shifted spectral feature is absent in the case of $C_{spiro}$–O bond cleavage, suggesting that it may serve as a fingerprint for photophysical relaxation via $C_{spiro}$–N bond breakage and reformation.

In what follows, the red-shifted nitrogen K-edge TRXAS signal associated with $C_{spiro}$–N bond dissociation, and the absence of a similar signal for $C_{spiro}$–O bond fission, are rationalized. The feature below 400 eV arises mainly from $S_{0/1} \to c_1$ transitions, where $c_1$ denotes the lowest-lying core-excited state. The analysis therefore focuses on how the energies of these transitions depend on the $C_{spiro}$–N and $C_{spiro}$–O bond lengths. Let us first consider the early-time dynamics, before $S_1 \to S_0$ internal conversion. At $t = 0$ fs, the $S_1 \to c_1$ transition is located at ~400 eV, as indicated by the red markers in Fig. 8, and has a negligible oscillator strength. As time progresses, this transition undergoes a red-shift with increasing $C_{spiro}$–N bond length and a blue-shift with elongation of the $C_{spiro}$–O bond. In both cases, these energetic shifts are accompanied by increases in oscillator strength, an effect more pronounced for $C_{spiro}$–N bond cleavage. To identify the origin of the opposing energetic shifts observed for the $S_1 \to c_1$ transition as a function of the $C_{spiro}$–N and $C_{spiro}$–O bond lengths, Fig. 9 shows the time-dependent valence and core-excited potential energy curves for the two SP1mt trajectories included in Fig. 8.

The corresponding results for two representative SP1mc trajectories are shown in Supplementary Fig. 24 and are qualitatively similar. Elongation of both the $C_{spiro}$–O and $C_{spiro}$–N bonds leads to a reduction in the $S_1$ state energy. The $c_1$ state energy, however, responds markedly differently to the two bond extensions: it is largely independent of the $C_{spiro}$–O bond length, while it exhibits a clear negative correlation with the $C_{spiro}$–N bond length. As a result, for $C_{spiro}$–O bond fission, the reduction in the $S_1$ state energy with increasing $C_{spiro}$–O bond length leads to a blue-shift of the $S_1 \to c_1$ transition. In contrast, for $C_{spiro}$–N bond fission, the decrease in the $S_1$ state energy is counterbalanced by a larger reduction in the $c_1$ state energy, with the net effect being a red-shift in the $S_1 \to c_1$ transition.

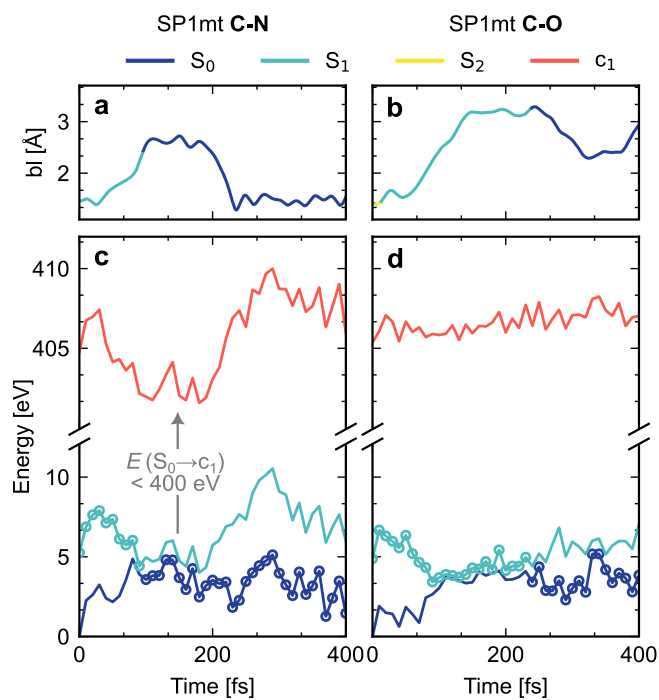

**Fig. 9 | Origin of the $C_{spiro}$–N bond-length-correlated red-shift in the nitrogen K-edge TRXAS spectra. c, d** Show time-dependent valence- and core-excited potential energy curves computed with RASSCF followed by RASSI every 10 fs along representative SP1mt surface hopping trajectories undergoing $C_{spiro}$–N and $C_{spiro}$–O bond fission, respectively. The circles indicate the active valence electronic states from the surface hopping simulations. The $S_2$ potential energy curves are excluded for clarity. **a, b** Show the time evolution of the bond lengths (bl) of the dissociating $C_{spiro}$–N and $C_{spiro}$–O bonds, respectively. The color coding of the bond lengths indicates the active states from the surface hopping simulations.

The elongations of the $C_{spiro}$–N and $C_{spiro}$–O bonds reduce the $S_1/S_0$ energy gaps, leading to $S_1 \to S_0$ internal conversions, as described in the preceding subsection. The electronic transitions themselves leave no signatures in the nitrogen K-edge TRXAS spectra. For $C_{spiro}$–N bond fission, the energy of the $c_1$ state increases when the $C_{spiro}$–N bond reforms on the $S_0$ state, returning to roughly its value at $t = 0$ fs. At the same time, the energy of the $S_0$ state decreases. These two effects act in the same direction, resulting in a pronounced blue-shift of the $S_0 \rightarrow c_1$ transition as the bond reforms.

The origin of the transient spectral feature below 400 eV in the nitrogen K-edge TRXAS spectra is thus the negative correlation between the energy of the $c_1$ state and the $C_{spiro}$–N bond length. A similar negative correlation between the energy of the lowest-lying core-excited state and a bond length was recently reported theoretically at the NEVPT2 level by Penfold and Curchod for cyclobutanone[58]. As noted therein, this trend is consistent with Natoli's rule[59], a semiempirical relationship between the absorption energy $E$ and the bond length $r$ according to $Er^2$ = constant.

A limitation of the present work is that, due to the considerable computational cost associated with X-ray spectroscopic calculations for a molecule of spiropyran's size and complexity, the TRXAS spectra were computed for four representative surface hopping trajectories rather than for the entire ensemble. Ensemble-level spectra are more directly comparable to experimental TRXAS data. At the ensemble level, the intensity of the TRXAS feature characteristic of $C_{spiro}$–N bond fission is expected to depend on the quantum yield of the photophysical relaxation. The quantum yield is, as discussed in the preceding subsection, challenging to predict theoretically, but the experimentally reported yields of at least 90%[6,8] are promising for detectability. The intensity is also expected to be affected by the degree of coherence of the $C_{spiro}$–N bond breakage and reformation. The coherence also affects the lifetime of the feature and, consequently, the experimental time resolution required to resolve it. The dynamics simulations reported in the preceding subsection predict large differences in this respect between the two ground state conformers (Fig. 5). The high degree of coherence exhibited by trajectories initiated from the SP1mc conformer is anticipated to yield an intense feature below 400 eV that lasts for ~150 fs. The larger spread in the time domain of the $C_{spiro}$–N bond fission and reformation for the SP1mt conformer is expected to result in a weaker signal persisting for several hundred femtoseconds. We note that the degree of coherence of the bond breaking and reformation may, like the quantum yield, be challenging to predict accurately theoretically with dynamics simulations; it may depend on the electronic structure method, the initial condition sampling technique, and possibly the spiropyran model. A comparison with a potential future TRXAS experiment could therefore provide valuable insights.

We conclude with some brief considerations on the pump wavelength for a potential future TRXAS experiment. In the present work, we simulated the TRXAS spectra following excitation to $S_1$. At the TDA-$\omega$B97XD level, the $S_1$ state lies at ~270 nm for SP3m (Supplementary Table 3), suggesting that it could be excited by the 3rd harmonic of an 800 nm laser (~266 nm). However, experimental UV-visible spectra recorded for SP3m in nonpolar solvents[5,51] suggest that higher-lying states than $S_1$ may be accessed upon 266 nm excitation. Should a pump wavelength of 266 nm be considered for an experiment, it would therefore be prudent to calculate spectra from higher-lying states beyond $S_1$, to ensure that core excitations from these states do not produce spectral features overlapping with that characteristic of $C_{spiro}$–N bond fission.

## Conclusions

In this work, we demonstrated theoretically that TRXAS at the nitrogen K-edge has the potential to fingerprint $C_{spiro}$–N bond breakage and reformation in spiropyran. The time evolution of the coupled electron-nuclear dynamics of spiropyran upon $S_1$ excitation was simulated using surface hopping coupled with TDA-$\omega$B97XD for the electronic structure, and nitrogen K-edge TRXAS spectra were calculated with the RAS approach at snapshots from representative trajectories. Consistent with the theoretically established picture of the photodynamics of spiropyran, the dynamics simulations predict competing $S_1 \rightarrow S_0$ relaxation pathways through two CIs, located along the $C_{spiro}$–O and $C_{spiro}$–N bond extensions, respectively. Relaxation through the former CI leads, for the large majority of the trajectories, to photoproduct formation. Relaxation through the latter is photophysical, since the $C_{spiro}$–N bond reforms on the electronic ground state. The TRXAS calculations indicate that the energy of the lowest-lying core-excited state exhibits a negative correlation with the $C_{spiro}$–N bond length. Photophysical relaxation via $C_{spiro}$–N bond breakage therefore results in a red-shifted $S_{1/0} \rightarrow c_1$ TRXAS signal that vanishes upon bond reformation on the ground state. This feature is absent for trajectories that follow the competing photochemical route since the energy of the lowest-lying nitrogen K-edge core-excited state is insensitive to the $C_{spiro}$–O bond length. This establishes the red-shifted feature as a fingerprint for photophysical relaxation. While caution must be exercised as the spectra were computed for a limited set of representative trajectories rather than for the full ensemble, our results indicate that nitrogen K-edge TRXAS may provide an experimental route to test the photophysical relaxation mechanism first proposed theoretically[11] more than a decade ago.

## Data availability

The structures and active states for all surface hopping trajectories are provided in Supplementary Data 1, and the optimized geometries of SP0mc, SP0mt, SP1mc, SP1mt, SP3mc, and SP3mt are provided in Supplementary Data 2. The remaining datasets generated during the current study are available from the corresponding authors on reasonable request.

### Acknowledgements
L.F. thanks Denis Jacquemin for suggesting the SP1m model system. This work received financial support under the EUR LUMOMAT project and the Investments for the Future program ANR-18-EURE-0012 (M.V. and L.F.) L.F. acknowledges thesis funding from the Région Pays de la Loire and Nantes University. B.R. acknowledges master internship funding from the Région Pays de la Loire through the Étoiles Montantes program. The project is also partly funded by the European Union through ERC Grant No. 101040356 (M. V.). The views and opinions expressed are however those of the authors only and do not necessarily reflect those of the European Union or the European Research Council Executive Agency. Neither the European Union nor the granting authority can be held responsible for them. The simulations in this work were performed using HPC resources from CCIPL (Le centre de calcul intensif des Pays de la Loire) and from GENCI-IDRIS (Grant 2021-101353).

### Author contributions
L.F., S.C., S.N. and M.V. designed the project. B.R. simulated the non-adiabatic dynamics under the supervision of L.F. and M.V. L.F. performed the TRXAS calculations with the RASSCF method under the supervision of M.V. S.C. performed the XAS calculations using the CCSD and IMOM-TDA approaches. L.F., B.R., S.C. and M.V. analysed the results. L.F. and M.V. wrote the manuscript. All authors contributed to the review of the manuscript.

### Competing interests
The authors declare no competing interests.

### Additional information
**Supplementary information** The online version contains supplementary material available at https://doi.org/10.1038/s42005-025-02175-1.

**Correspondence** and requests for materials should be addressed to Lina Fransén or Morgane Vacher.

**Peer review information** *Communications Physics* thanks the anonymous reviewers for their contribution to the peer review of this work. A peer review file is available.

**Reprints and permissions information** is available at http://www.nature.com/reprints

**Publisher's note** Springer Nature remains neutral with regard to jurisdictional claims in published maps and institutional affiliations.